# Interfacial area transport for bubbly-to-slug transition flows in small diameter pipes

Zhuoran Dang[a,∗], Mamoru Ishii[a,∗]

[a]*School of Nuclear Engineering, Purdue University, 516 Northwestern Ave., West Lafayette, IN, 47907, USA*

**Abstract**

This study aims to experimentally investigate the two-group interfacial area transport in small diameter pipes. Experimental data focusing on the bubbly to slug transition regime, namely one-group to two-group transport region, are collected in a 12.7 mm vertical pipe under adiabatic, air-water conditions. The result shows the intergroup transfer in the small diameter pipe can be drastic, especially under low superficial liquid velocities. The cause of this phenomenon is mainly due to the large relative bubble size comparing to the pipe cross-sectional area. The wake entrainment effect could be enhanced by the small spherical bubbles that are acting like cap or slug bubbles in a medium-size pipe. Based on the experimental observation, a throughout analysis of the dependence of the drastic intergroup transfer is provided in this study. The models predicting the initiation of drastic intergroup transfer in small diameter pipes in terms of the bubble diameter and the void fraction are developed. These models are compared with the two-phase data among the different pipe sizes and the results show a good agreement. These newly developed models are applied to the IATE wake entrainment model by developing a transition function analogous to the sigmoid function. With the transition function, the revised IATE model is given the new ability on predicting the drastic intergroup transfer phenomenon.

∗Corresponding authors
*Email addresses:* zdang@purdue.edu (Zhuoran Dang), ishii@purdue.edu (Mamoru Ishii)

**Nomenclature**

| | |
|---|---|
| $a$ | Relaxation parameter [−] |
| $a_i$ | Interfacial area concentration (IAC) [$m^{-1}$] |
| $C_D$ | Drag coefficient [−] |
| $D_c$ | Maximum distorted bubble diameter [$m$] |
| $D_{d,max}$ | Maximum stable bubble diameter [$m$] |
| $D_h$ | Channel hydraulic diameter [$m$] |
| $D_m$ | Bubble motion diameter [$m$] |
| $D_{re}$ | Critical bubble diameter by Reynolds number [$m$] |
| $D_{sm}$ | Bubble Sauter mean diameter [$m$] |
| $D_{tr}$ | Critical bubble diameter by geometry effect [$m$] |
| $f_{os}$ | Bubble oscillation frequency [$Hz$] |
| $g$ | Gravitational acceleration [$m/s^2$] |
| $h$ | Height [$m$] |
| $j$ | Superficial velocity [$m/s$] |
| $L_w$ | Wake region length [$m$] |
| $L_w^*$ | Dimensionless wake region length [−] |
| $Re$ | Reynolds number [−] |
| $Sr$ | Strouhal number [−] |
| $v$ | Bubble velocity [$m/s$] |
| $V$ | Bubble volume [$m^3$] |
| $We$ | Weber number [−] |

*Greek letters*

| | |
|---|---|
| $\alpha$ | Void fraction [−] |
| $\beta$ | relaxation parameter [−] |
| $\lambda$ | Flow channel area factor [−] |
| $\mu$ | Kinematic viscosity [$m^2/s$] |
| $\psi$ | Bubble shape factor [−] |
| $\rho$ | Density [$kg/m^3$] |
| $\sigma$ | Surface tension [*N/m*] |
| $\theta_b$ | Inclination angle [*rad*] |
| $\theta_w$ | Cap or slug wake angle [*rad*] |

*Subscripts*

| | |
|---|---|
| $b$ | bubble |
| $e$ | equivalent |
| *eff* | effective |
| *exp* | experiment |
| $f$ | liquid |
| $g$ | gas |
| $0$ | atmospheric |
| *WE* | wake entrainment |
| *SO* | shearing-off |

*Mathematical symbols*

| | |
|---|---|
| $<\ >$ | area-averaged quantity |
| $<<\ >>$ | void fraction weighted-mean quantity |

# 1. Introduction

The dynamic prediction of interfacial area concentration (IAC), a key parameter of phase interaction terms in the two-fluid model, has been realized by the interfacial area transport equation (IATE) [1]. The advantage of the method is that it is inherently consistent with the change mechanism of the interfacial area. The IAC can be predicted free of static flow regime dependencies, therefore, avoiding the discontinuities caused by the transition between flow regimes. The state-of-art two-group IATE is capable to give predictions to comprehensive two-phase flow conditions from bubbly to churn-turbulent flow regimes [2]. However, recent studies [3, 4] reported the performance of the IATE on the transition flows is not satisfying. In the conventional approach of IATE model development [5, 6, 7], the establishment of mechanistic models of the bubble interactions is based on the understanding of bubble transport observed through experimental observations. In this sense, rigorous and comprehensive experimental studies on the interfacial area transport in the flow regime transitions are needed for the development of the IATE.

The interfacial structure and bubble dynamics of the two-phase flow depend on many features such as properties of the fluids, flow channel geometry and length, and the design of two-phase mixture injector, etc. In this sense, the understanding and modeling of these topics should base on a wide range of different experimental setups, while the patterns should be observed and summarized based on similar experimental setups. For example, to investigate the pipe diameter effect on the interfacial structure, the experimental setups that are used to generate databases should be as similar as possible except for the pipe diameter. The previous studies [5, 6, 9, 10, 30] with different pipe diameters but similar injecting systems can be used along with the current study. To investigate the effect of bubble injector design, experimental data under similar pipe sizes should be used. [31, 32]

The pipe size effect on the interfacial structure and bubble dynamics of the two-phase flow has been qualitative discussed. [30, 33] In a large diameter pipe, a slug bubble that covers the whole flow channel cross-sectional area usually does not exist due to the bubble surface instability. Whereas in small diameter pipes, stable slug bubbles are usually observed. The existence of the bubbles that could cover the whole flow channel area substantially changes the two-phase flow structure and dynamics. A large or small diameter pipe is usually determined by comparing the pipe diameter with the maximum stable bubble diameter, which is defined by Kotaoka and Ishii [8] based on Taylor instability,

$$D_{d,\max} = 40\sqrt{\frac{\sigma}{g\Delta\rho}} \tag{1}$$

In the development of the IATE, one important subject is the model developed in the one-group to two-group interfacial area transport, which corresponds to the bubbly-to-slug flow transition in small diameter pipes, or bubbly-to-cap-bubbly flow transition in large diameter pipes. Specifically, in small diameter pipes, bubble transport could be affected by the containing wall. Worosz [4] pointed out the enhanced wake entrainment effect in bubbly-to-cap-bubbly transition flows in a 50.8 mm inner diameter pipe. Recent experimental studies [9, 10] on interfacial area transport in a 25.4 mm inner diameter pipe observed that the void fraction and IAC transport between the two groups could be drastic in bubbly-to-slug transition flows. This drastic intergroup transport phenomenon is beyond the predicting capability of the state-of-art two-group IATE. Wang et al. [10] explained this unique phenomenon using the relatively large bubble size that increases the intensity of the bubble coalescence due to the wake entrainment. To fully understand this phenomenon, this study performed air-water two-phase flow experiments on a 12.7 mm vertical upward pipe in the bubbly-to-slug transition flows. An experimental study on this flow channel geometry and focusing on the current flow conditions has not been performed in the previous literature [11]. The experimental analysis focuses on demonstrating the developing interfacial structures along the flow direction due to the bubble interactions, as well as focusing on the intergroup transfers. Based on the available database, this study proposes a group of models that predicts the onset of the drastic intergroup transfer, and these models serve on extending the predicting capability of the IATE.

## 2. Experimental

### 2.1. Experimental facility

The test facility in the present study uses the same test facility as the previous study [12], as shown in Fig. 1. The test section is made of transparent acrylic pipes with its inner diameter and length of 12.7 mm and 4117 mm, respectively. A diagram figure describing the injection system in the test facility is given in Fig. 2. Air is supplied by a compressor and enters into the test section through a porous media pipe with its pore size of 40 $\mu m$. Water is supplied by a centrifugal pump and enters into the test section by dividing it into a primary and secondary flow. The primary water flow controls the uniformity of the injected two-phase flow and the secondary water flow determines the inlet air bubble size. The air and water flow rates are measured using rotameters and magnetic flow meters, respectively. Local two-phase flow structures are measured using four-sensor conductivity probes at three axial locations of $z/D_h$ = 31, 156, 282, and 10 radial positions from $r/R$ = 0 to 0.875. To quantify the gas expansion, local pressure measurements are conducted using differential pressure transducers. The selection of the flow conditions is based on the flow inlet boundary conditions including superficial liquid and gas velocity $<j_f>$ and $<j_{g,0}>$, and the void fraction $\alpha$. The subscript $0$ refers that $<j_{g,0}>$ is calculated under the atmospheric pressure. To investigate the characteristics of interfacial area transport in bubbly to slug transition flows and address the drastic intergroup transfer phenomenon, 23 flow conditions are selected based on the Mishima-Ishii flow regime map [13], plotted in Fig. 3. Table 1 gives the quantitative information of these flow conditions.

Four-sensor conductivity probe [14, 15] is used for the measurement of two-phase flow parameters, including time-averaged void fraction, IAC, and bubble interfacial velocity. The principle of the four-sensor conductivity probe in measuring these two-phase parameters are based on the previous studies [14, 16, 17, 18], and the detailed information could refer to the previous research [12, 19]. As reported by Kim et al. [15], The measurement uncertainty of the four-sensor conductivity probe in measuring the IAC is less than ±10%. Since this study focuses on the two-group interfacial area transport, the bubbles measured by the four-sensor probe are classified into two groups based on the bubble size. the boundary of the two groups is the maximum distorted bubble diameter $D_c$, expressed as,

$$D_c = 4\sqrt{\frac{\sigma}{g\Delta\rho}} \quad (2)$$

The two-group bubbles are classified during the experimental signal processing, by comparing the measured chord length of each bubble with the maximum distorted bubble diameter. Then the time-averaged void fraction, interfacial area concentration, and bubble velocity for each group are calculated by averaging the bubbles in each group. This method of using bubble chord length is considered to be a valid approximation approach for classifying the bubbles. This is because, under the current experimental setup, the sizes of the two-group bubbles are quite different from each other and are not near the maximum distorted bubble size. Therefore, the error that a Group-2 bubble is misclassified into a Group-1 bubble by using the bubble chord length as the approximate value of bubble diameter is considered to be negligibly small. In consist with the previous studies on two-group bubble theory, bubbles with their sizes smaller than $D_c$, namely small spherical or elliptical bubbles, are called Group-1 bubbles, and large, distorted bubbles larger than $D_c$ are called Group-2 bubbles. The difference in bubble size and shape leads to substantial differences in transport phenomena due to the differences in drag force and liquid-bubble interaction mechanisms. [20] The bubble size is quantified using bubble Sauter mean diameter, calculated using the following equation,

$$D_{sm} = 6\frac{\alpha}{a_i} \qquad (3)$$

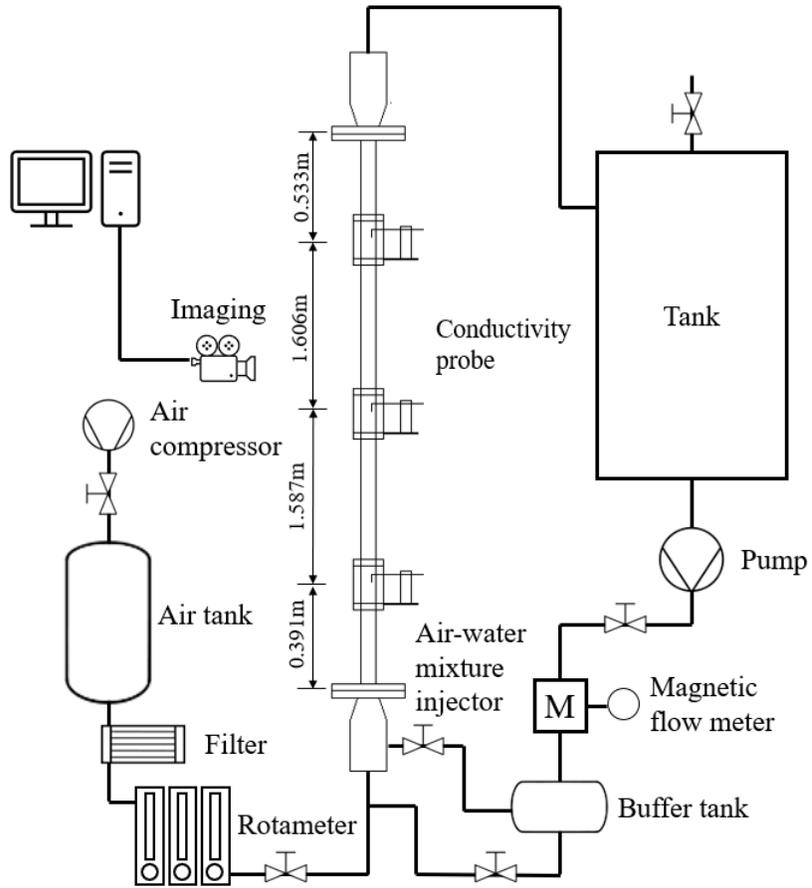

Figure 1: Schematic of the experimental facility.

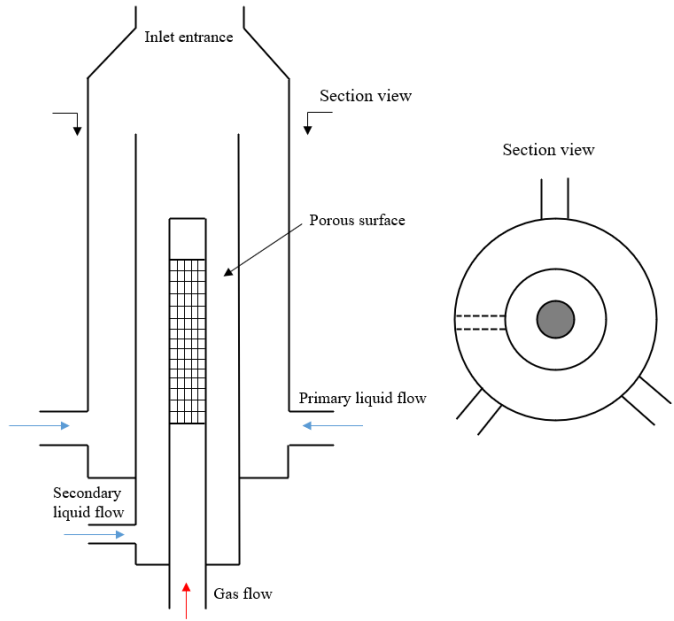

Figure 2: Two-phase mixing and injecting system.

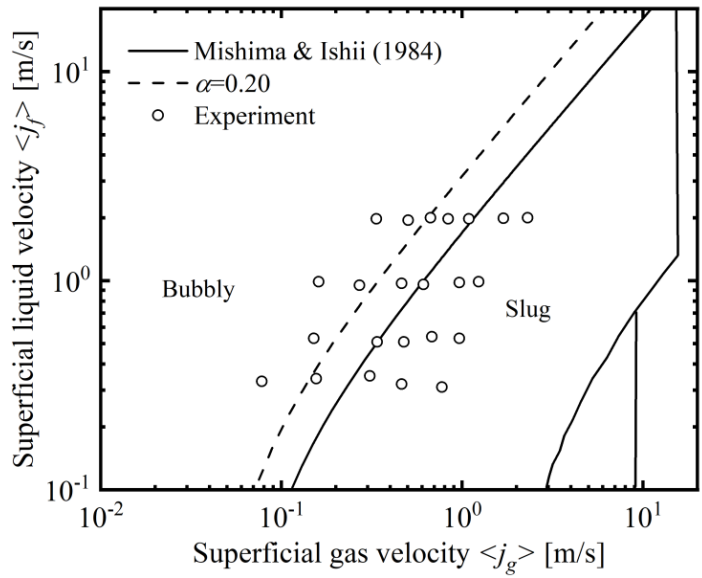

Figure 3: Test matrix.

Table 1: Flow conditions.

| $<j_f>$ [m/s] | $<j_{g,0}>$ [m/s]; $\alpha_{z/D=31}$ [−] | $P_{z/D=31}$ [kPa]; $dP/dz$ [kPa/m] | $<j_f>$ [m/s] | $<j_{g,0}>$ [m/s]; $\alpha_{z/D=31}$ [−] | $P_{z/D=31}$ [kPa]; $dP/dz$ [kPa/m] |
|---|---|---|---|---|---|
| 0.3 | **0.078**(0.122) | **124.020**(5.036) | 0.5 | **0.151**(0.154) | **134.711**(7.930) |
|  | **0.156**(0.221) | **130.308**(6.684) |  | **0.339**(0.301) | **136.291**(8.372) |
|  | **0.308**(0.356) | **131.709**(7.056) |  | **0.475**(0.391) | **136.781**(8.498) |
|  | **0.462**(0.390) | **135.67**(8.096) |  | **0.679**(0.532) | **137.890**(8.790) |
|  | **0.772**(0.469) | **137.962**(8.720) |  | **0.966**(0.603) | **140.966**(9.422) |
| 1.0 | **0.161**(0.098) | **135.091**(8.006) | 2.0 | **0.335**(0.081) | **151.064**(11.773) |
|  | **0.270**(0.170) | **137.278**(8.512) |  | **0.502**(0.124) | **151.996**(12.386) |
|  | **0.463**(0.228) | **140.372**(9.319) |  | **0.669**(0.214) | **152.747**(12.326) |
|  | **0.610**(0.346) | **141.208**(9.548) |  | **0.837**(0.210) | **153.771**(12.972) |
|  | **0.966**(0.447) | **143.158**(10.053) |  | **1.088**(0.258) | **155.016**(13.348) |
|  | **1.235**(0.462) | **148.429**(12.211) |  | **1.689**(0.338) | **156.492**(13.764) |
|  |  |  |  | **2.301**(0.379) | **156.913**(13.580) |

## 3. Results and discussion

*3.1. Experimental result of the two-phase profile*

The radial and axial profiles of the two-phase parameters are presented and discussed in this section. In Fig. 4, the results of three flow conditions with a similar superficial gas velocity $<j_{g,0}> \approx 0.15 m/s$ are plotted, and each row contains the results from the same flow condition. The time-averaged radial profiles including void fractions, IACs, and bubble velocities are presented that each parameter is plotted in the same column. The results collected in different axial positions can be found in the same sub-figure. The results in this figure clearly show how the flow boundary conditions can affect the radial distributions of the two-phase parameters and the interfacial area transport. At the measurement position $z/D = 31$ that is close to the inlet of the test section, the void fraction and IAC distribution can shift from a nearly uniform distribution to wall-peak distribution with the increase of liquid flow rate. As the flow develops, the distributions change to core-peaked. It indicates that small bubbles become larger and migrate to the center due to the net transverse forces including shear-induced lift force and wall lubrication force, as mentioned in the previous studies [4, 9, 12]. The increase of bubble size along the flow direction can be caused by three factors [21]: small bubble expanses due to the pressure change, the advection effect caused by the bubble velocity change, and the bubble interactions. From the bubble velocity result, the aforementioned advection effect does not attribute to the increase of bubble void fraction or bubble size in these flow conditions as the bubble velocity increases along the flow direction. After the radial distribution becomes core-peaked, the intergroup transfer can happen, as shown in the flow condition with $<j_f> = 0.3 m/s$. In this flow condition, the Group-2 bubbles are formed with the appearance and increase of the Group-2 void fractions and IACs at the following measurement positions $z/D = 156, 282$. At the same time, the Group-1 shares correspondingly decrease. The degree of this intergroup transfer is quite high that neither it appears in the experimental results on a medium-size pipe, nor it is predictable using the current versions of IATE.

Fig. 5 shows the radial profiles of three test conditions under the same superficial liquid velocity $<j_f> \approx 0.5 m/s$, the plotting rules of which correspond to those in Fig. 4. By comparing these flow conditions, the drastic intergroup transfer can be observed more clearly. As $<j_g>$ increases, the initial void fraction and bubble size are getting larger, thus the void and IAC distributions at $z/D = 31$ change from wall-peaked to core-peaked. These two flow conditions with larger $<j_g>$s show the drastic intergroup transfer phenomenon. It indicates that the drastic intergroup transfer phenomenon happens because the change of flow boundary conditions ($<jf>$ and $<jg>$) leads to the change of void fraction and bubble size, resulting in the change of the flow structure and bubble interactions.

To further analyze the interfacial area transport and the intergroup transfer that happens in these flow conditions, the corresponding axial development of the two-phase parameters are presented in Fig. 6 and 7. In the figures, the area-averaged two-phase parameters including the two-group bubble Sauter mean diameter are presented. From the results, when the drastic intergroup transfer happens, the void fractions experience a large amount of transfer between groups that the profiles of the void fractions show an "X" shape. The Group-1 IAC drops sharply, while Group-2 IAC raises relatively slow due to the large size of the Group-2 bubbles. As a result, the total IAC drops at a high rate during the drastic intergroup transfer. From the bubble Sauter mean diameter $D_{sm}$ result, the Group-1 bubble size increases slowly before the drastic intergroup transfer. When the $D_{sm}$ reaches certain values around 4 mm-4.5mm, the drastic intergroup transfer can be initiated. During the drastic intergroup transfer, the Group-1 $D_{sm}$ decreases since the large Group-1 bubbles are transferring into Group-2 bubbles and it reduces the mean $D_{sm}$ value. To better identify the effect of two-phase flow boundary conditions on the intergroup transfer, void fraction and IAC profile maps are plotted in Fig. 8 and 9, corresponding to the flow conditions plotted on the Mishima-Ishii flow regime map in Fig. 3. From the maps, it can be seen that the drastic intergroup transfer can happen in $<j_f>$ and moderate $<j_g>$ flow conditions. Increasing $<j_f>$ can raise the $<j_g>$ needed for the drastic intergroup transfer to be initiated. The drastic intergroup transfer does not happen in high $<j_f> = 2.0 m/s$.

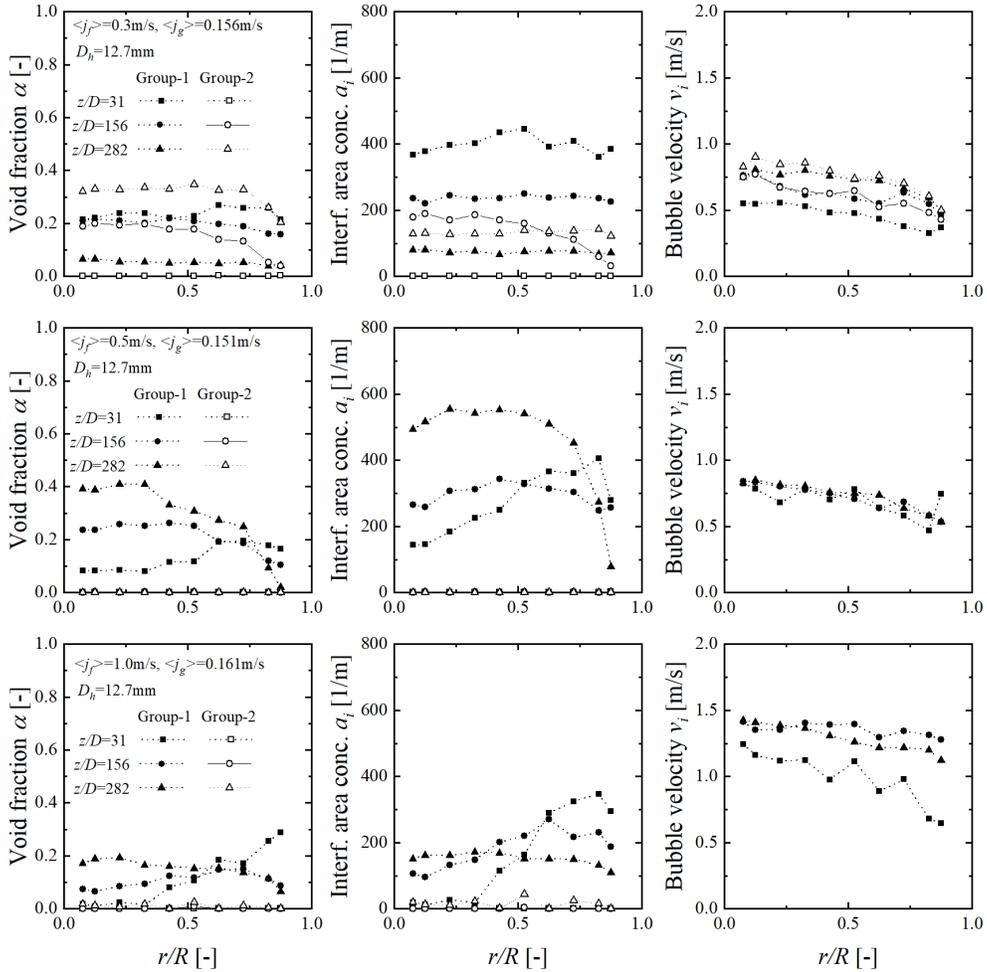

Figure 4: Two-group time-averaged interfacial parameters of three flow conditions with a similar $<j_g> \approx 0.15 m/s$.

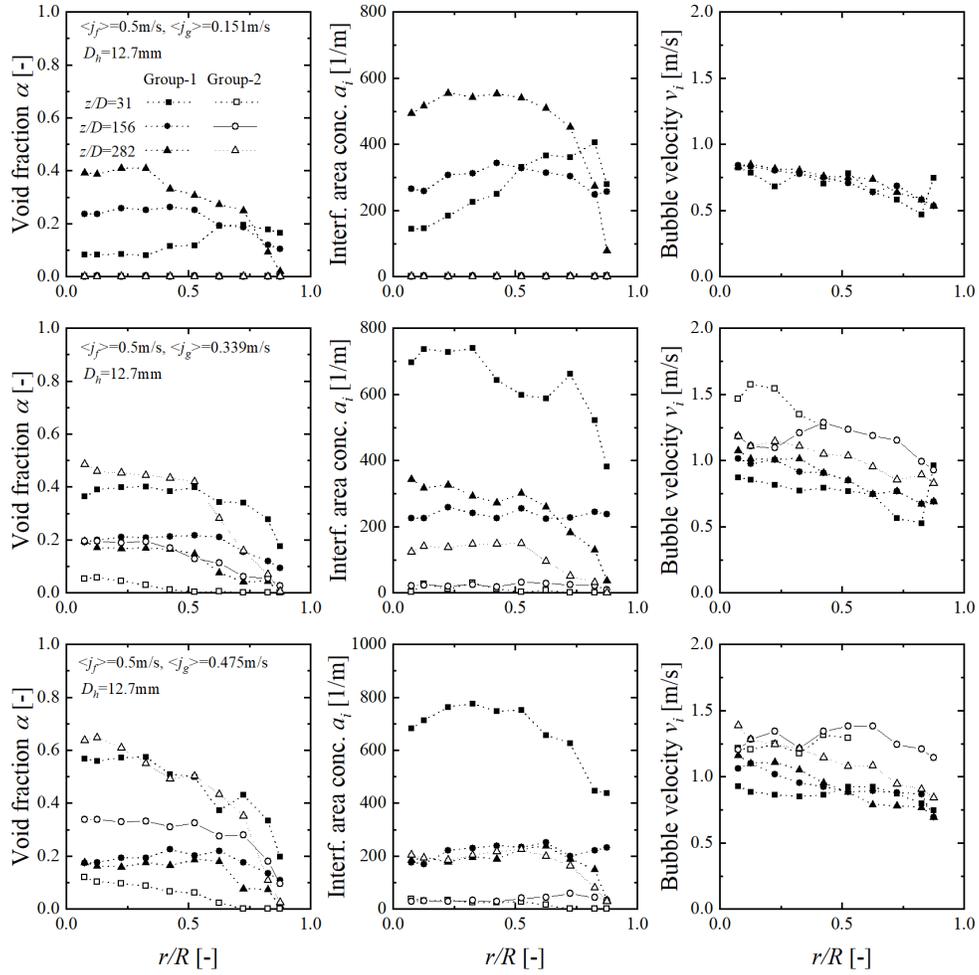

Figure 5: Two-group time-averaged interfacial parameters of three flow conditions with a similar $\langle j_f \rangle \approx 0.5 m/s$.

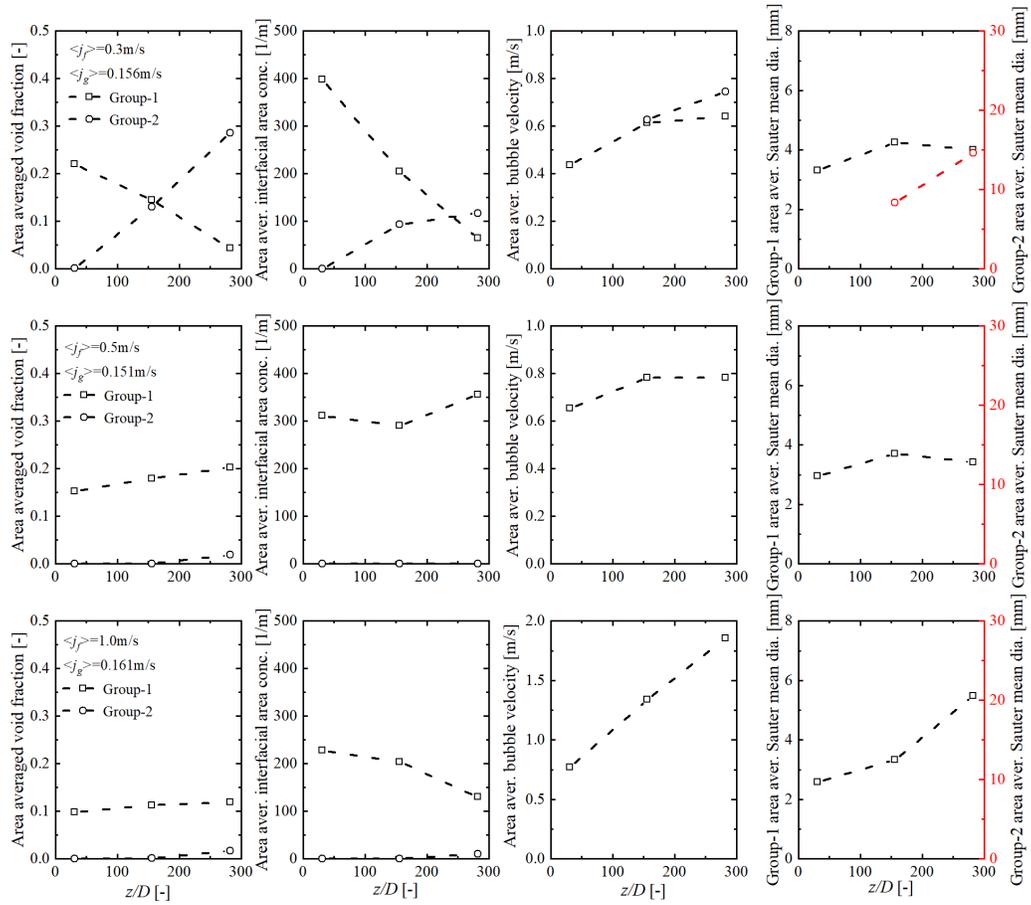

Figure 6: Two-group area-averaged interfacial parameters of three flow conditions with a similar $\langle j_g \rangle \approx 0.15 m/s$.

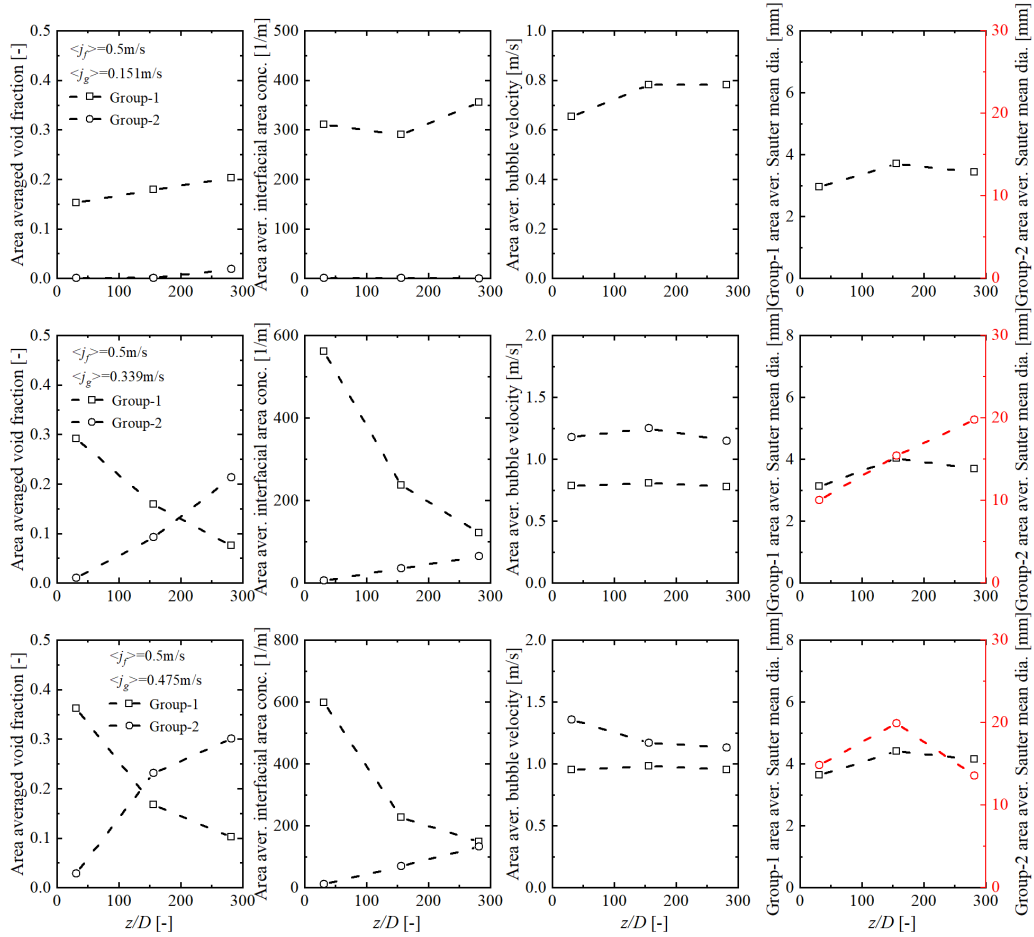

Figure 7: Two-group area-averaged interfacial parameters of three flow conditions with a similar $\langle j_f \rangle \approx 0.5 m/s$.

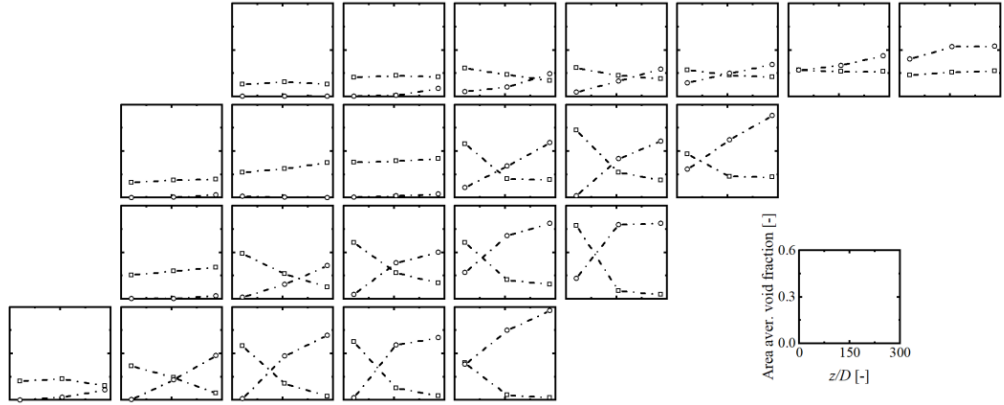

Figure 8: Two-group area-averaged void fraction profile map.

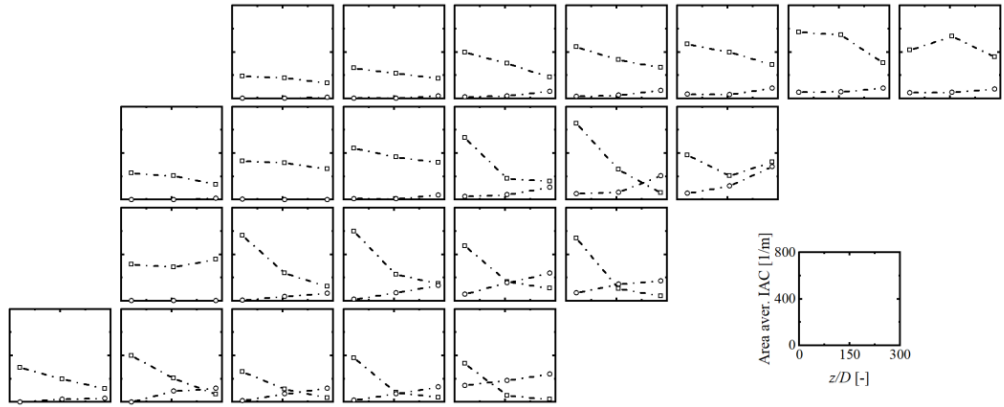

Figure 9: Two-group area-averaged IAC profile map.

### 3.2. Initiation of the drastic intergroup transfer

In the previous discussion of the experimental results, the occurrence of the drastic intergroup transfer is related to the two-phase flow boundary conditions (e.g., $j_f$, $j_g$, etc.). These boundary conditions influence the development of the two-phase flow structure by altering the flow structure and bubble interactions.

Essentially, the drastic intergroup transfer phenomenon can be regarded as an intensive shift of flow structure, which can be regarded as the combination of phases with various topographic features [22]. The change of the flow structure could be due to the change in the percentages of each phase or the change of the topological shape. Therefore, the two-phase parameters that directly describes these changes should be used to determine the change of the flow structure. Specifically, void fraction and bubble size are two of the two-phase topological parameters. They are strongly related to the boundary conditions and also could be experimentally measured. One approach is to estimate the critical quantities of these parameters at which the drastic intergroup transfer could be initiated. In this section, a theoretical analysis of the effects of the bubble size and the void fraction that affect the initiation of the drastic intergroup transfer is presented.

*3.2.1. Effect of bubble size*

In the IATE [1], the small bubble coalescence mechanisms are classified as the random collision that bubbles collide and coalesce driven by turbulence and the wake entrainment that the following bubbles are entrained by the preceding bubbles and coalesce. [5] It has been proved that in small diameter pipes, the wake entrainment mechanism can be dominant and lead to the shift from one-group to two-group interfacial area transport. [4] Therefore, one approach to analyzing the bubble size effect, as well as the void fraction effect, on the intergroup transfer is to associate it with the wake entrainment mechanism.

The wake region of a leading bubble in small diameter pipes can easily cover the whole flow channel cross-section than that in large diameter pipes. Due to the constraint from the containing wall, a leading bubble can have more trailing bubbles included in its wake region in small pipes than in large pipes. Thus, the leading bubble can be more "effective" in small pipes. The method that models this "effectiveness" of the leading bubble in a certain size of the flow channel is to find the boundary situation by comparing the wake region cross-sectional area and the flow channel cross-sectional area. To realize it, the bubble rising path or rising motion in the flow should be considered first.

As a bubble rises through the water, it is not only moving in the rising direction but also moving in either spiral or zigzag path, namely bubble secondary motion [23]. Fan and Tsuchiya [23] pointed out that this is because the energy generated through bubble rising in a low viscosity medium requires to be dissipated by additional motions. Therefore, this energy induces the bubble to oscillate, known as bubble secondary flow motion. To simply this theory, it is assumed that a Group-1 bubble moves in a spiral motion and the projected area covered by this moving bubble can be described using a diameter called motion diameter, $D_m$. It is further assumed that the amplitude of developed bubble zigzag motion can be used to determine the spiral motion diameter, which is proved experimentally. [24] This concept is depicted in Figure 10. Based on the above setup, the motion diameter can be expressed by the partition of drift velocity $v_{gj}$ in the radial direction and the spiral motion frequency,

$$D_m \approx D_b + \frac{v_{gj} \sin(\theta_b)}{2 f_{os}} \tag{4}$$

where $\theta_b$ and $f_{os}$ are the inclination angle of the bubble and the bubble oscillation frequency, respectively. From Miyahara et al. [25], $\theta_b$ is almost constant (~25°) when the bubble Reynolds number is less than 3000. It then decreases rapidly when the Reynolds number increases. The oscillation frequency ($f_{os}$) is usually represented by the Strouhal number ($Sr_b = f_{os} D_b / v_b$). Tsuge and Hibino [26] analyzed their data based on dimensional analysis and found a unique relationship between $Sr_b$ and the drag coefficient $C_D$, expressed by the following correlation,

$$Sr_b = \begin{cases} 0.100 C_D^{0.734} & (C_D \leq 2) \\ 6.13 \times 10^{-3} C_D^{4.71} & (C_D > 2) \end{cases} \tag{5}$$

Therefore, the relation between the bubble diameter and the bubble motion diameter is expressed as

$$D_m = D_b \left(1 + \frac{v_{gj} \sin(\theta_b)}{2 Sr_b v_b}\right) \tag{6}$$

Now consider the onset of the significance of an effective leading bubble. The objective is to find a geometrical relationship between the diameter of the effective leading bubble and the flow channel hydraulic diameter. An extreme/boundary situation that all the bubbles with a certain diameter are considered to be effective leading bubbles is considered. Similar to the approach modeling the bubble coalescence used by the study of Mishima and Ishii [13], these moving bubbles are in an arrangement similar to triangle shape so that they can cover almost all the cross-sectional area of the flow channel, as depicted in Figure 10. In this case, the relation between the bubble diameter and the flow channel hydraulic diameter is expressed as

$$\frac{D_m}{D_h} = \left(\frac{3}{2\sqrt{3}+3}\right)\lambda \tag{7}$$

where $\lambda$ is the percentage of the area that bubbles can occupy in the flow cross-section. In the study of Fu et al. [27], $\lambda$ is approximately equal to 0.9 when considering the whole flow channel is occupied by a fully developed slug bubble, indicating the maximum space that the void can reach. Whereas Clift et al. [28] stated that $\lambda$ can be as small as 0.6. Therefore, $\lambda$ should be a value between 0.6 and 0.9. Based on the above equations, the critical bubble diameter can be calculated as,

$$D_{tr} = \frac{0.464\lambda}{1+\frac{v_{gj}\sin(\theta_b)}{2Sr_b v_b}} D_h \tag{8}$$

From the above equation, $D_{tr}$ is related to the pipe size. However, one important notion is that the bubble should also be large enough to form an effective wake region behind it. This can be quantified using the Bubble Reynolds number $Re(D_{re})$, which is proportional to the wake area/length formed behind a bubble or sphere in a continuous stream. Figure 11 shows the experimental bubble Sauter mean diameters against the critical bubble diameters that determine the initiation of drastic intergroup transfer, arranging based on $j_f$. The experimental results of 12.7 mm (left) and 25.4 mm (right) inner diameter pipe [10] match satisfyingly with the predictions of the critical bubble diameters.

*3.2.2. Effect of void fraction*

Besides the bubble diameter, another parameter that influences the wake entrainment interaction in small diameter pipes is the void fraction. Consider the area under the effect of a bubble wake region. Ideally, if the wake region is large enough to cover the whole flow channel, then the wake entrainment mechanism happens in a good possibility since if more gas is added into the flow channel, it will be in the wake region. This situation corresponds to a critical void fraction that can be estimated through the relation between the gas bubble and the wake region it creates.

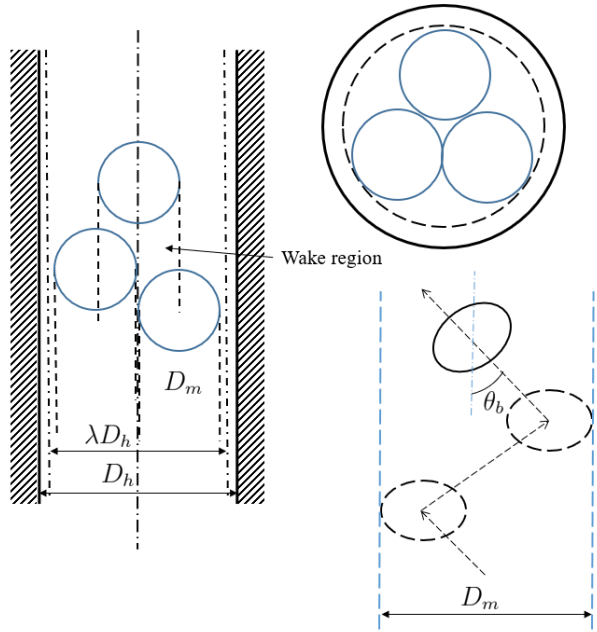

Figure 10: Conceptual diagram showing the relation between the bubble motion diameter and the pipe size.

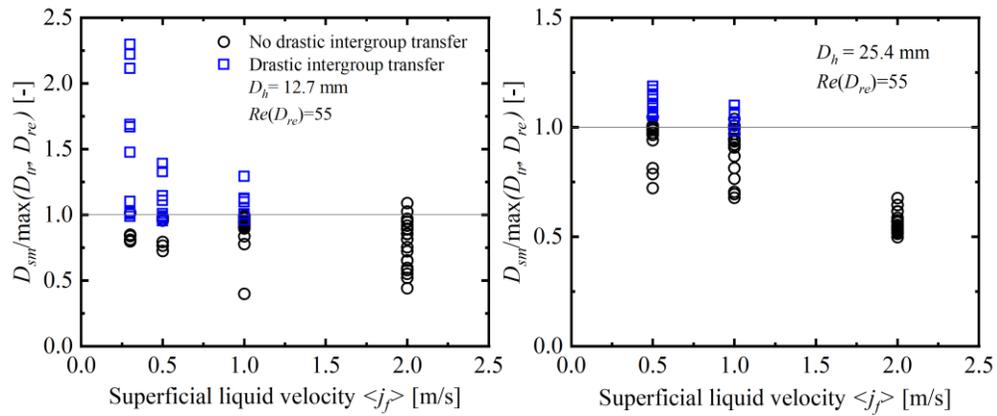

Figure 11: The relation of the ratio of bubble diameter and critical diameters and the existence of drastic intergroup transfer: Experimental results at all measurement positions in the current study (12.7 mm) and Wang et al (25.4mm) [10]

Assuming that initially before the large irregular bubble is formed, all the bubble in the flow channel is small and spherical. The spherical bubble volume is calculated as follows,

$$V_b = \frac{1}{6}\pi D_b^3 \qquad (9)$$

Assuming the bubble wake region can be approximated to cylindrical geometry. The total volume of the bubble and the wake region it forms is expressed as follows,

$$V_{WE} = \frac{1}{12}\pi D_b^3 + \frac{1}{4}\pi D_b^2 L_w \qquad (10)$$

where $L_w$ is the length of the wake region and it is related to the bubble diameter. According to the study by Fan and Tsuchiya [23], the wake length is roughly 5-7 times the bubble diameter in an air-water system [5]. The wake length can then be expressed into a dimensionless form,

$$L_w^* = \frac{L_w}{D_b} \approx [5, 7] \qquad (11)$$

Thus, Eq.10 can be transformed as,

$$V_{WE} = \frac{1}{12}\pi D_b^3 (1 + 3L_w^*) \qquad (12)$$

Consider that the wake region only exists behind the bubbles. It means that the wake region has no effect on the preceding bubbles and only works on the following bubbles. Thus, to model the relation between the gas volume and the effective wake region volume, the wake region volume $V_{WE}$ should be approximated to be divided by 2, which could be expressed as follows,

$$\frac{V_{WE,eff}}{V_b} = \frac{V_{WE}/2}{V_b} = \frac{(1 + 3L_w^*)}{4} \doteq [4, 5.5] \qquad (13)$$

where $V_{WE,eff}$ is the effective volume of wake region. The relation described in Eq. 13 could be further transformed by introducing the void fraction. By definition, gas void fraction $\alpha = V_b/V$ and wake region volume fraction $\alpha_{WE} = V_{WE,eff}/V$, where $V$ denotes the total volume of the flow channel. Based on these definitions, the relation between the void fraction and the wake volume fraction could be derived based on Eq. 13,

$$\alpha_{WE} = [4, 5.5]\alpha \qquad (14)$$

If $α_{WE}$ equals one, which means that ideally, the wake entrainment volume can cover the whole flow channel if perfectly distributed without overlapping. In this case, the corresponding void fraction could be derived as,

$$0.182 < α_{tr} < 0.25 \tag{15}$$

The subscript *tr* denotes the critical value, corresponding to the subscript of bubble diameter in Eq. 8. Eq. 15 means that ideally when the averaged void fraction is above the values, the wake entrainment phenomenon starts. However, the modeling process assumes that the position of the void should be ideally distributed so that all the flow channel area should be covered, which in reality cannot happen. When the averaged void fraction is around $α_{tr}$, some wake entrained areas can overlap each other, and some other areas can stay unaffected. However, due to the nature of air-water flows, at this level of void fraction, although bubbles are not uniformly distributed, they are rather small so that the wake entrainment effect is small. Therefore, the αtr can be regarded as a valid approximation.

Based on the above discussion, the critical void fraction that marks when the wake entrainment effect happens is derived. the average void fraction is estimated to be $α_{tr} = 0.216$. It matches the experimental observations in both 12.7 mm and 25.4 mm ID pipe [10], as depicted in Figure 13. However, since the drastic intergroup transfer phenomenon is not observed in high superficial liquid velocity. The critical void fraction should not be solely used as the criterion. The models that describe the initiation of the drastic intergroup transfer phenomenon should be formulated based on both critical void fraction $α_{tr}$, the critical bubble diameters $D_{tr}$ and $D_{re}$.

*3.2.3. Transition function*

An intergroup transfer model that helps improve the prediction of IATE can be developed based on the above models. In the previous section, the critical parameters are derived to quantify the onset of the drastic intergroup transfer. To give the IATE the ability to predict this phenomenon, a continuous function formulated based on these models should be developed. Similar to the study of Worosz [4], a transition function that is analogous to the Sigmoid function ($S(x) = 1/(1 + e^{-x})$) is proposed,

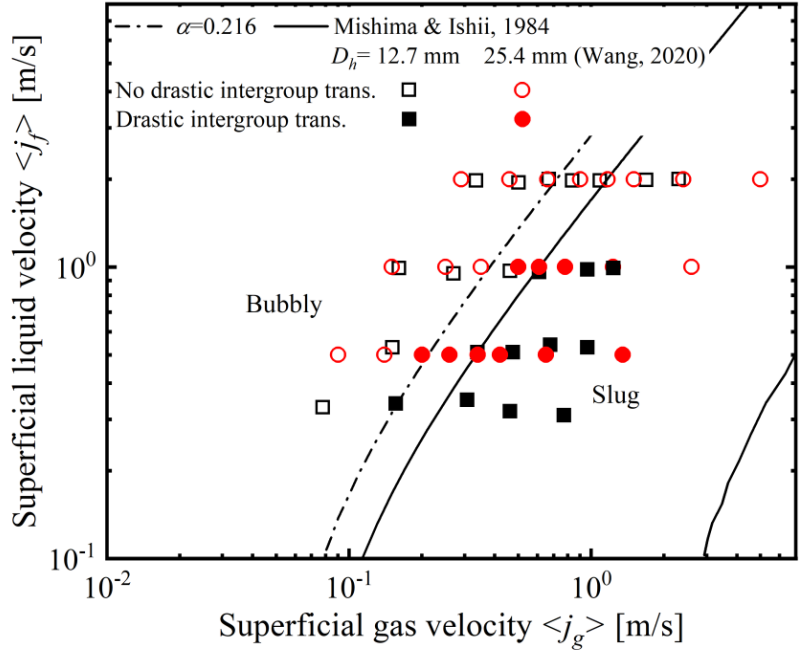

Figure 12: Experimental observations identifying the appearance of the drastic intergroup change and the critical void fraction boundary.

The function $T(D_{sm},\alpha)$ is described as:

$$T(D_{sm}, \alpha) = a + \frac{(1-2a)}{(1+e^{-g(D_{sm})})(1+e^{-h(\alpha)})}$$
$$g(D_{sm}) = \frac{(D_{sm} - max[D_{tr}, D_{re}])}{\beta D_c}$$
$$h(\alpha) = \frac{(\alpha - \alpha_{tr})}{\beta} \tag{16}$$

where $a$ and $\beta$ are relaxation parameters determining the range and the slope of the transition function, and representing the significance of the drastic intergroup transfer and the credibility of the models, respectively. The values of these two parameters are determined through the experiments. Fig. 13 shows an example of the transition function when $max[D_{tr},D_{re}] = 4mm$ and $\alpha$ is greater than $\alpha_{tr}$. In the model prediction for real cases, these parameters are calculated based on the boundary conditions.

The transition function is designed to be applied to the models of bubble coalescing rate due to wake entrainment by simply being multiplied with the current models,

$$R_{WE}^{(11,2)} \rightarrow R_{WE}^{(11,2)} T(D_{sm}, \alpha)/a \qquad (17)$$

The above equation is valid when the Group-1 bubble size is assumed to be the same, which is utilized in Worosz's model [4]. However, if the bubble size is not assumed to be the same and the bubble number density distribution is specified, Eq. 16 could also be applied to the leading bubbles, whose number density can be estimated using the bubble number density distribution function. The performance of IATE with the new wake entrainment model Eq. 17 is given in Fig. 14. The bubble interaction terms are based on the study of Worosz [4] that specifically focuses on the one-group to two-group transition flows in a 50.8 mm ID pipe. In the current IATE model, only the rate of Group-1 bubble coalescing and creating Group-2 bubbles due to wake entrainment ($WE$11,2) is modified as given in Eq. 17. Besides the new model, some experiment-determined coefficients related to the Group-2 bubble are modified based on the extended databases of the 12.7 mm and 25.4 mm ID pipes: $C_{WE}^{(22,2)} = 0.15 \rightarrow 0.4; C_{SO}^{(2,12)} = 0.12 \rightarrow 0.002$. The predictions of the original Worosz model are also provided for comparison. The results show that by applying the transition function, the revised model not only keeps its good predictions as the previous model does but also gives improved prediction results when the drastic intergroup transfer happens.

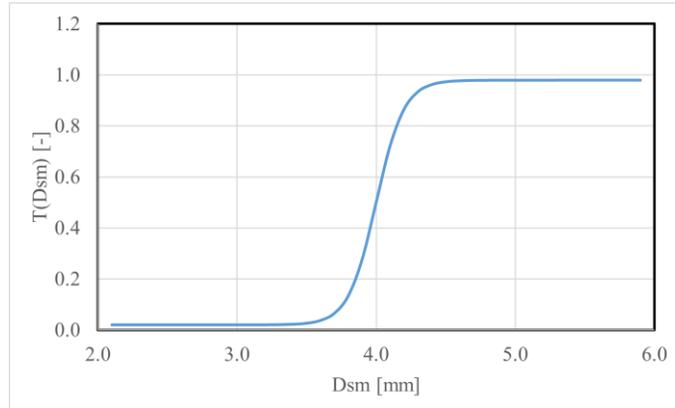

Figure 13: Example of transition function $T(D_{sm})$ with $a = 0.02$, $max[D_{tr}, D_{re}] = 4mm$, and $\beta$ = 9e-5. $\alpha$ is set to be greater than $\alpha_{tr}$ and equals to 0.25.

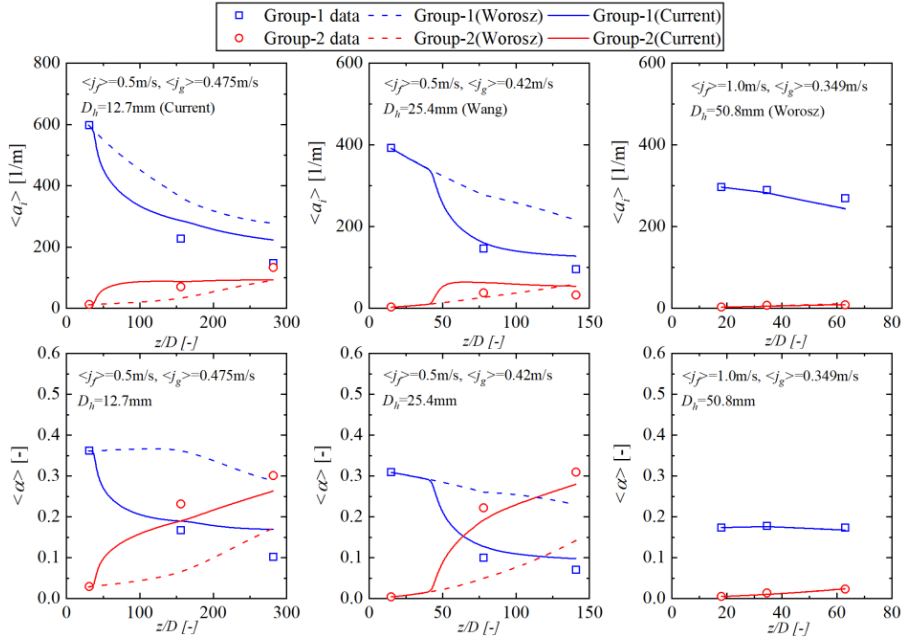

Figure 14: Predictions of the original Worosz model and the new model with the transition function $T(D_{sm}, \alpha)$ ($a = 0.01; \beta = $ 9e-6) against the experimental data on 12.7 mm, 25.4 mm [10] and 50.8 mm [4] ID pipes.

## 4. Conclusion

In this study, two-group interfacial area transport in small diameter pipes are investigated based on the experiments. It focuses on the bubbly-to-slug (one-group to two-group) transition flows in small diameter pipes, and the experiments are performed in a 12.7 mm ID pipe under air-water, adiabatic condition using four-sensor electrical conductivity probes. The interfacial area transport characteristics are presented and discussed based on the two-phase parameters including void fraction, IAC, bubble velocity, and bubble Sauter mean diameter. From the results, the intergroup transfer in the small diameter pipe can be drastic, especially under low superficial liquid velocities $<j_f>$. The reason that this phenomenon can happen in small diameter pipes is due to the large relative bubble size to the pipe cross-sectional area. The wake entrainment effect could be enhanced by these spherical or elliptical bubbles that are acting like cap or slug bubbles in a medium-size pipe. From the experiment, the averaged bubble size and void fraction should reach certain values before the intergroup transfer becomes drastic. Therefore, a theoretical analysis of the effect of bubble size and void fraction on the drastic intergroup transfer is performed and models of the initiation of this phenomenon ($α_{tr}$, $D_{tr}$, and $D_{re}$) are proposed. These newly developed models are applied to the IATE wake entrainment model by developing a transition function analogous to the sigmoid function. With the transition function, the revised IATE model is given the new ability on predicting the drastic intergroup transfer phenomenon.